\documentclass[nologo,11pt,a4paper]{ETHpaper}           

\usepackage[pdftex]{graphicx}
\usepackage{subfigure}
\usepackage{amsmath}     
\usepackage{color, colortbl}
\usepackage{afterpage}
\usepackage{url}
\usepackage{multirow}
\usepackage{rotating}
\usepackage{hyperref}
\definecolor{Gray}{rgb}{0.9,0.9,0.9}
 
\begin{document}
  
\title{Categorizing Bugs with Social Networks: A Case Study on Four Open Source Software Communities}
\titlealternative{Categorizing Bugs with Social Networks: A Case Study on Four Open Source Software Communities}
\author{Marcelo Serrano Zanetti, Ingo Scholtes,\\ Claudio Juan Tessone and Frank Schweitzer}
\authoralternative{Marcelo Serrano Zanetti, Ingo Scholtes, Claudio Juan Tessone and Frank Schweitzer}
\address{Chair of Systems Design, ETH Zurich, Switzerland\\
  \url{www.sg.ethz.ch}} 
\reference{to appear in the proceedings of the 35th International Conference on Software Engineering (ICSE 2013) - Software Engineering in Practice (SEIP) Track.}
\www{\url{http://www.sg.ethz.ch}} 
\makeframing
\maketitle 
 
\begin{abstract}
Efficient bug triaging procedures are an important precondition for successful collaborative software engineering projects. 
Triaging bugs can become a laborious task particularly in open source software (OSS) projects with a large base of comparably inexperienced part-time contributors.
In this paper, we propose an efficient and practical method to identify \emph{valid} bug reports which a) refer to an actual software bug, b) are not duplicates and c) contain enough information to be processed right away. Our classification is based on nine measures to quantify the social embeddedness of bug reporters in the collaboration network. 
We demonstrate its applicability in a case study, using a comprehensive data set of more than $700,000$ bug reports obtained from the \textsc{Bugzilla} installation of four major OSS communities, for a period of more than ten years.
For those projects that exhibit the lowest fraction of valid bug reports, we find that the bug reporters' position in the collaboration network is a strong indicator for the quality of bug reports.
Based on this finding, we develop an automated classification scheme that can easily be integrated into bug tracking platforms and analyze its performance in the considered OSS communities. A support vector machine (SVM) to identify \emph{valid} bug reports based on the nine measures yields a precision of up to $90.3 \%$ with an associated recall of $38.9 \%$. 
With this, we significantly improve the results obtained in previous case studies for an automated early identification of bugs that are eventually fixed.
Furthermore, our study highlights the potential of using quantitative measures of social organization in collaborative software engineering.
It also opens a broad perspective for the integration of social awareness in the design of support infrastructures.
\end{abstract}

\section{Introduction\label{sec:intro}}


Triaging and processing bug reports is an important task in collaborative software engineering which can crucially affect product quality, project reputation, user motivation and thus the long-term success of a project.
Practical experience from large open source software (OSS) projects confirms that --particularly in projects with large numbers of comparably inexperienced part-time contributors-- the process of triaging, categorizing and prioritizing bug reports can become a laborious and difficult task that consumes considerable resources.
Both the importance and complexity of this problem can be illustrated by a simple example:
Out of the more than $64,000$ bug reports that have been resolved by the community of the \textsc{Mozilla Firefox} project, more than $50,000$ (or $\approx78\%$) of these reports have eventually been identified either as \emph{duplicates} of known bugs, \emph{invalid} reports that refer to a user error rather than a software issue or \emph{incomplete} reports which lack basic information required to reproduce the alleged bug.
The magnitude of this problem in large-scale projects calls for (semi-)automated techniques that assist bug handling communities in the triaging and prioritization of bug reports.
The provision of methods which are able to automatically identify \emph{valid} bug reports with high precision can have huge implications for practitioners of distributed software engineering: Being able to filter, assign and prioritize those bug reports that likely result in a bug fix can significantly improve the responsiveness of support communities.
Furthermore, a temporary deferral of those bug reports that are likely to be duplicates, invalid or incomplete to a moderation queue can considerably alleviate the effort required for bug triaging.
It can also be used to automatically enforce the adherence to community guidelines, e.g.~by automatically asking original reporters to reconfirm that reported bugs are neither duplicates nor incomplete.

Due to the importance for practical software engineering, a number of different approaches for the automated classification of bug reports have been studied, among them approaches based on the automated assessment of information provided by bug reports \cite{Anvik2006,Hooimeijer2007,Bhattacharya2010,Shihab2010}, natural language processing \cite{Runeson2007,Wang2008,Cubrani2004}, the temporal dynamics of bug handling processes \cite{Podgurski2003}, coordination patterns \cite{cataldo2008}, or the reputation of bug reporters \cite{Guo2010,Xuan2012,Zimmermann2012}.
Based on a unique data set containing the full history of more than $700,000$ bug reports in four major OSS communities, in this paper we consider to what extent automated bug classification techniques can be based on \emph{quantitative measures for the social embeddedness of bug reporters in the project's community}.
We particularly address this question from the perspective of complex, evolving collaboration networks and the computation of node-centric metrics that capture structural properties like centrality and clustering.

Our contributions to the current state of research are the following:

\begin{itemize}
    \item We study relations between the centrality of bug reporters and the eventual outcome of the bug triaging process. For the four OSS communities studied in this paper, we find strong evidence for the hypothesis that the centrality of users in the collaboration network is indicative for the quality of bug reports.
    \item We show that quantitative measures for the bug reporter's position in the collaboration network can be used for an automated classification of valid bug reports. For the four studied OSS communities, we find that this classification achieves a precision of up to $90.3 \%$ with an associated recall of $38.9 \%$.
\end{itemize}

With this, we extend previous works that have studied automated classification of bugs that are eventually fixed.
In particular, we use a more comprehensive data set, more sophisticated quantitative measures for user's position in the evolving structures of a community as well as a predictive modeling approach that is based on a support vector machine.
In the following section, we provide a more detailed review of existing literature on automated bug classification and prediction mechanism as well social aspects of collaborative software engineering. From this we then extract a set of open research questions that are addressed in the remaining sections of this paper.

\section{Social Aspects in Bug Report Processing\label{sec:related}}

The distribution of contributions, the structure and evolution of collaboration networks in OSS projects, as well as their relation with individual and collective performance have been studied in a number of works.
A quantitative study of the development efforts in the projects \textsc{Apache} and \textsc{Mozilla} has been presented in \cite{mockus2002}.
Among other aspects, the distribution of contributions across community members has been analyzed.
For the \textsc{Apache} project, the authors particularly validate that - while coding efforts are mainly concentrated on a small set of core developers - the bug handling and reporting process is based on a much larger community of part-time contributors.

Apart from the mere distribution of contributions, the topology of communication and collaborations between contributors is an interesting field of study.
The relation between the network position of developers in bug handling communities and their success rate (in terms of the number of bugs the developers fix) has been studied in \cite{Ehrlich2012}.
There, the authors find that developers with higher node degree fix bugs at a higher rate.
Furthermore the authors provide implications for future research, calling for subsequent studies of the relation between communication structures and individual as well as team-based performance.
Our work complements the study of \cite{Ehrlich2012} in the sense that we investigate the relation between the centrality of bug reporters and their individual performance, i.e. whether the reports are eventually found to refer to actual software issues.
Our methods are based on earlier work quantifying the dynamics of social organization in OSS communities \cite{zanettiICCSW2012}.
Social mechanisms underlying the impact of communication topologies on bug handling performance have been studied in \cite{Bettenburg2008}.
There, the authors conclude that the most difficult task of successfully handling bugs is the mediation between the users and the developers of a project.
Similar results have been presented by the authors of \cite{Wang2011}, whose analysis is based on the bug handling communities of two major OSS projects.
Their analysis verifies that the collaborative identification of the cause of a software defect is one of the most difficult tasks that needs to be solved before bugs can be properly addressed by developers.
Based on data obtained from the \textsc{Bugzilla} community of the \textsc{Eclipse} project and similar to our approach, in \cite{Bettenburg2010} measures of communication dynamics and user centrality have been studied in networks constructed based on user comments and \emph{CC} subscriptions.
The findings suggest that the centrality of users in the communication flow networks extracted from \textsc{Bugzilla} data is related to the future failure proneness of code.
Similarly, the relationship between communication structures and success at the collective level has been studied in \cite{Wolf2009} and \cite{Wolf2009a}.
In those papers, the use of social network structures and communication deficiencies for the prediction of build failures has been proposed.
Furthermore, it was found that positive team performance is related to communication structures that facilitate information dissemination.
These quantitative insights about the social dimension of software engineering highlight the importance of social indicators and provide an important foundation for our approach of using related measures from social network analysis for the classification of bug report quality.

Due to the difficulty of handling user contributed bug reports in large-scale projects with millions of users, a number of different approaches for supporting bug triaging processes based on an automatic classification of bug reports have been studied.
In \cite{Hooimeijer2007} a simple linear regression model for the quality of bug reports has been proposed based on a data set of $27,984$ bug reports from the project \textsc{Mozilla Firefox}.
The model is based both on information available at the time of submission as well as post-submission data like the number of comments or attachments added during the first hours and days.
The evaluation of a model based on this data shows that there is a $5 \%$ increase of predictive power compared to a pure chance prediction.
In a case study on the \textsc{Eclipse} project \cite{Shihab2010}, a predictive model has been introduced that is based on the textual information in comments and the bug description.
The analysis shows that the model yields a precision of $62.9 \%$ and a recall of $84.5 \%$ when predicting which bugs will be reopened after being marked as closed.
Apart from simple regression models, machine learning approaches have been used for the automated classification and triaging of bug reports in a number of works \cite{Lucca2002,Anvik2006,Bhattacharya2010,Podgurski2003,Somasundaram2012}.
In \cite{Anvik2006}, the use of machine learning techniques for assisting humans in assigning bugs to developers has been proposed.
In \cite{Bhattacharya2010} a machine learning approach is used to reduce bug tossing, i.e. the simultaneous assignment of bugs to multiple developers.
The authors show that bug tossing can be reduced significantly when classifying developers according to the product relationships of previously fixed bugs.
In \cite{Somasundaram2012} different machine learning approaches have been applied to bug descriptions and comments stored in the \textsc{Bugzilla} database of the \textsc{Eclipse} project.
Here the authors prove the suitability of support vector machines and Latent Dirichlet Allocations for the prediction of the category of bug reports.

Indicators for the \emph{social context} of users have been considered for the prediction of which bugs get fixed and which are likely to be reopened in \cite{Guo2010,Zimmermann2012}.
In \cite{Guo2010}, a number of bug report features have been used, including the reputation of bug reporters in terms of the fraction of their previously reported bugs that were eventually fixed.
The authors show that a statistical model for the automated identification of those bugs that will get fixed can yield a precision of $68 \%$ and a recall of $64 \%$.
The same approach has recently shown to be successful for the prediction which bugs get reopened \cite{Zimmermann2012}.

Data from the \textsc{Bugzilla} installations of \textsc{Eclipse} and \textsc{Mozilla} have been used in \cite{Xuan2012} to model developer prioritization in bug repositories.
Here the authors used a ranking of developers based on social networks and apply a support vector machine to predict the severity of bug reports assigned to developers.
In \cite{graphbasedpamela2012}, a predictive model for the bug severity based on the location of the defect in the software dependency network has been studied.
Here the authors find that the degree of components in the software is indicative for the severity of bugs. 
  
\section{Study Design and Methodology}
\label{sec:methods}

Based on a the review of existing work that is related to a) the influence of social embeddedness on the performance of communities and individual contributors and b) the automated classification of bug reports, we identify the following open research questions which will be addressed in our paper:

\begin{itemize}
\item [\textbf RQ1] Is the position of bug reporters in the evolving collaboration structures of bug handling communities related to the quality of contributed bug reports?
\item [\textbf RQ2] Can quantitative measures for the position of bug reporters be used to predict which bug reports refer to valid bugs?
\end{itemize}

With the prediction methodology proposed in section \ref{sec:classify}, we extend and improve previous approaches to automated bug classification in a number of ways:
First we consider a larger data set which contains a total of more than $5.8$ million time-stamped change events for more than $700,000$ bug reports from four large OSS projects.
Second, rather than using simple, one-dimensional social indicators like the number of previously submitted reports or the number of connections, we use a set of nine topological measures to quantify the position of bug reporters in the collaboration network, among them a comprehensive set of centrality measures as well as degree, local clustering structure and membership in the largest network component.
Third, rather than taking a simple static perspective, we consider \emph{evolving collaboration networks} by using fine-grained temporal data on collaboration and communication events.
Based on these features, we apply a machine learning approach for predicting which of the bug reports are eventually identified as valid, i.e. which are referring to actual bugs that need to be addressed by the community.
We further strictly limit our prediction methodology to \emph{only include information available at the time of the submission of a bug report, thus making the approach directly applicable in a practical setting.}
To the best of our knowledge, no prior work has combined such a comprehensive set of network measures on evolving networks with a machine learning classifier and applied it to data set of similar scale.
Our findings show that our methods significantly improve the precision and recall of existing automated bug classification schemes.



In our paper, we adopt a data-driven approach that is based on a data set we collected from the \textsc{Mozilla Bugzilla}\cite{serranobugzilla2005} installations of the four communities evolving around the following OSS projects: \textsc{Mozilla Firefox}, \textsc{Mozilla Thunderbird}, \textsc{Eclipse} and \textsc{NetBeans}.
In the following, we provide a detailed description of a) the data retrieval process and the categories of bug reports available in the data, b) our methodology of extracting time-stamped collaboration networks and c) the measures applied in our analysis. 

\subsection{Data Retrieval\label{sec:data}}

Records retrievable via the \textsc{Bugzilla} \emph{API} are centered around \emph{bug reports} which are identified by a unique \emph{bug Id}. Further, users registered in the \textsc{Bugzilla} installation of the respective OSS project are also identified by their unique \emph{user Id}.
Each bug report has a number of associated fields, for which the history of all updates along with a time stamp and the \emph{Id} of the user who has changed the field, is stored in the database. For our analysis, we use the \emph{user Id} of the user who initially submitted the bug report (throughout the paper we will refer to this user as the \emph{bug reporter}), the time stamp of the initial submission, and the status of the bug report (like e.g. \emph{unconfirmed}, \emph{pending}, \emph{reproduced}, \emph{resolved}). We further use the \emph{user Id} of the so-called \emph{ASSIGNEE},  who is a user responsible for providing a fix to the bug, and a list of \emph{user Id's} of those users that have (or were) subscribed to receive subsequent updates on the bug report, \emph{CC}.

For our study, we retrieved the full history of all bug reports via the \emph{API} of the respective projects.
Our data set contains  roughly $715,000$ bug reports and $5.8$ Million change events recorded in the time between January 1999 and June 2012.
Table \ref{tab:TableStatistics} presents some basic statistics of the data set used throughout this paper.

\begin{table*}[!t]
\centering
\caption{Time periods, number of bugs, number of change events and number of bugs with particular status. The different bug resolution categories are the following: \emph{FIX} for fixed, \emph{DUP} for duplicate, \emph{INV} for invalid, \emph{WOF} for won't fix and finally \emph{INC} for incomplete. More details in section \ref{sec:data}.}
\scalebox{0.8}{
\begin{tabular}{|l|r|r|r|r|r|}
\label{tab:TableStatistics}
                                                        & \textsc{Firefox}           & \textsc{Thunderbird}   & \textsc{Eclipse}       & \textsc{NetBeans}      & Total \\
\hline
 \rowcolor{Gray}    Start date                          & April 2002        & January 2000  & October 2001  & January 1999  & $-$ \\
                    Total bug reports                   & 112,968           &  35,388       & 356,415       & 210,921       & 715,692\\
 \rowcolor{Gray}    Change events                       & 1,068,070         &  313,957             & 2,594,385     & 1,875,878     & 5,852,290 \\
                    Changes / report                    & 9.45              &     8.87          & 7.28          & 8.89          & 8.18 \\
 \rowcolor{Gray}    Resolved bugs (resolved/total)      &  64,088 (0.57)     &  21,644 (0.61)      & 158,957 (0.45)&  42,851 (0.19) &   287,540 (0.40) \\
\hline
 \rowcolor{Gray}    FIX (FIX / resolved)                &  10,856 (0.17)     &  4,508 (0.21) & 103,453 (0.65)&  21,442 (0.50) &   140,259 (0.49) \\
                    DUP (DUP / resolved)                &  24,263 (0.38)     & 10,336 (0.48) & 28,227 (0.18) &  9,328 (0.22)  &   72,154 (0.25) \\
 \rowcolor{Gray}    INV (INV /resolved)                 &  11,785 (0.18)     &  2,829 (0.13) & 12,601 (0.08) &  4,082 (0.10)  &   31,297 (0.11) \\
                    WOF (WOF / resolved)                &  2,708 (0.04)      &  581 (0.03)   & 14,676 (0.09) &  5,515 (0.13)  &   23,480 (0.08) \\
 \rowcolor{Gray}    INC (INC / resolved)                &   14,476 (0.23)     &  3,390 (0.16) & -  &2484 (0.06)  &    20,350 (0.07)
\end{tabular}}
\end{table*}

In particular, our analysis is focused on a subset of those $287,540$ bug reports that had a final status indicating that they were \emph{resolved}.
We limit our analysis to these bug reports because the bug handling community already completed the triaging process and thus reached a decision on how they were processed.
For this subset of resolved bugs we extract the full history of change events and categorize each bug according to the final change in the \emph{Resolution} field of the corresponding record.
Bugs that had a final \emph{Resolution} status of \emph{FIXED} (i.e. a bug fix has been created by a developer), \emph{INVALID} (i.e. the report refers to expected behavior or wrong usage rather than to a software bug), \emph{DUPLICATE} (i.e. the report refers to a bug that has already been reported) or \emph{WONTFIX} (i.e. the bug is valid and reproducible but it will not be fixed due to a lack of resources or low priority) were categorized accordingly.
In addition, we consider a bug report to fall into the category \emph{INCOMPLETE} whenever it had an intermediate status that indicates that the initial bug report was missing information required to properly triage the bug.
While the projects \textsc{Mozilla Firefox}, \textsc{Mozilla Thunderbird} and \textsc{NetBeans} make use of a specific status for incomplete reports, in the \textsc{Eclipse} community, bug reports that lack necessary information simply remain in the initial status \emph{NEW}.
Since this procedure does not allow us to easily classify corresponding bugs, we disregard the \emph{INCOMPLETE} category for the \textsc{Eclipse} project.

\subsection{Network Construction\label{sec:nets}}
Our approach to utilize measures for the embeddedness of users in the community is based on the extraction of social networks.
Those can be viewed as proxies for the collaboration and communication structure of an OSS project during a particular period of time.
Our data set is comprehensive in that it contains a history of all events associated with all bugs reported during a period of more than ten years.
For the construction of social networks we focus on those update events that directly capture dyadic interactions, and therefore can arguably be interpreted as pairwise interactions between users.
In particular, we use the dyadic relations \emph{ASSIGN} and \emph{CC} for this purpose.
For the present study, we decided to neglect additionally available information like the sequence of comments on bugs for which the inference of direct interactions between users is more difficult and necessarily error-prone.
Any user can add usernames to the \emph{CC} list of a bug report, which will make sure that the added user receives information on all future updated of a particular bug.
Special permissions are required for a user to \emph{ASSIGN} a bug to another user, which is hence being made responsible for providing a solution for the issue.
We would like to emphasise that focusing on \emph{CC} and \emph{ASSIGN} updates necessarily provides a limited perspective on the interactions between users.
Nevertheless we argue that the generated social networks are insightful with respect to the collaboration and communication structures of a project:
A \emph{CC} interaction between users $A$ and $B$ indicates that $A$ is aware of $B$ and that $A$ knows what $B$ is interested in.
In addition, an \emph{ASSIGN} interaction between $A$ and $B$ is indicative for different roles in the community. For example, user $A$ identifies the cause of a bug and assigns it to user  $B$ who is a developer and likely be able to fix it.

The simplest, and usually adopted, approach to analyze social networks in OSS communities is to study the topology by aggregating all interactions throughout the history of a project.
However, since our data set covers interactions from  more than one decade, the meaningfulness of such aggregated structures is questionable.
It is likely that most of the users represented by nodes in the aggregated network never have been active within the same time period.
This clearly limits the expressiveness of the network structure in terms of a project's ``social organization''.
In order to overcome this shortcoming, we make use of the fact that - like all other updates in our data set - \emph{CC} and \emph{ASSIGN} interactions have a precise time stamp.
In our analysis, we particularly study networks of collaborations constructed by aggregating all interactions occurring within time windows with a length of $30$ days.
This allows us to focus on collaboration networks existing at short periods of time during the project's history, e.g. when particular users were present, particular bugs were reported or when the project had a particular level of popularity and maturity.
In the following, we provide a detailed description of the quantitative measures used in our analysis of the resulting time-stamped collaboration networks.

\subsection{Network Measures\label{sec:measures}}

The literature is rich in measures to quantify structural features of (social) networks \cite{stanleysocial1994,newmannetsbook2010}. We adopt some of these measures to  capture the social organization in bug processing communities.

\subsubsection{Centrality measures\label{sec:cenmea}}

Node-centric measures of \emph{centrality} allow us to assess the relative \emph{importance} of nodes in a given network.
This importance, or centrality, can be expressed through different approaches.
The simplest one is the number of connections a node  has to other nodes,  known as the \emph{degree centrality}.
In a social context, degree centrality can be interpreted either in terms of the potential impact of a node on other nodes or as the amount of information available to a node.
However, degree centrality does not capture the actual \emph{position} of a node in the network in terms of how close an node  is to \emph{all} other nodes.
A further important measure is thus the so-called \emph{closeness centrality} \cite{freemancentrality1979}, which is defined as the inverse of the sum of all distances to all other nodes. The centrality of nodes can be also measured in terms of the role they  play in connecting other nodes.  The so-called \emph{betweenness centrality} is given by the total number of shortest paths between all possible pairs of nodes that pass through a node \cite{stanleysocial1994}.

\emph{Eigenvector centrality} is a more sophisticated feedback centrality measure in which the centrality of a node is recursively influenced by the centrality of its direct neighbors:
\begin{equation*}
Ev(n_i)=\frac{1}{\lambda}\sum_{n_j\in M(n_i)}Ev(n_j)
\end{equation*}
Here $M(n_i)$ is the set of direct neighbors of node $n_i$ and $\lambda$ is the largest eigenvalue of the network's adjacency matrix $A$ \cite{newmannetsbook2010}.
In other words, nodes connected to highly central nodes increase their own centrality.
For our analysis, we use the eigenvector centrality implementation of the \textsc{igraph} library \cite{igraph2006} for the \textsc{R} language \cite{R2012}.
The last two measures considered are the \emph{clustering coefficient} and \emph{k-coreness}.
The first captures to what degree two nodes that have a neighbor in common are also neighbors.
The second one is based on a network decomposition such that nodes are assigned to so-called \emph{shells} of the network topology.
Nodes belong to a given shell $k$ if they have a degree $k$ after removing all other nodes with degree up to $k-1$. Nodes in shells with higher number can be seen to have higher relative influence within a community \cite{Havlin2012}.

\subsubsection{Analysis of Largest Connected Component}

\begin{figure*}[!ht]
\begin{center}
\subfigure[\textsc{Eclipse} (Dec. 2002) 244 nodes, 319 links \label{fig:net_eclipse}]{
    \includegraphics[width=0.43\textwidth]{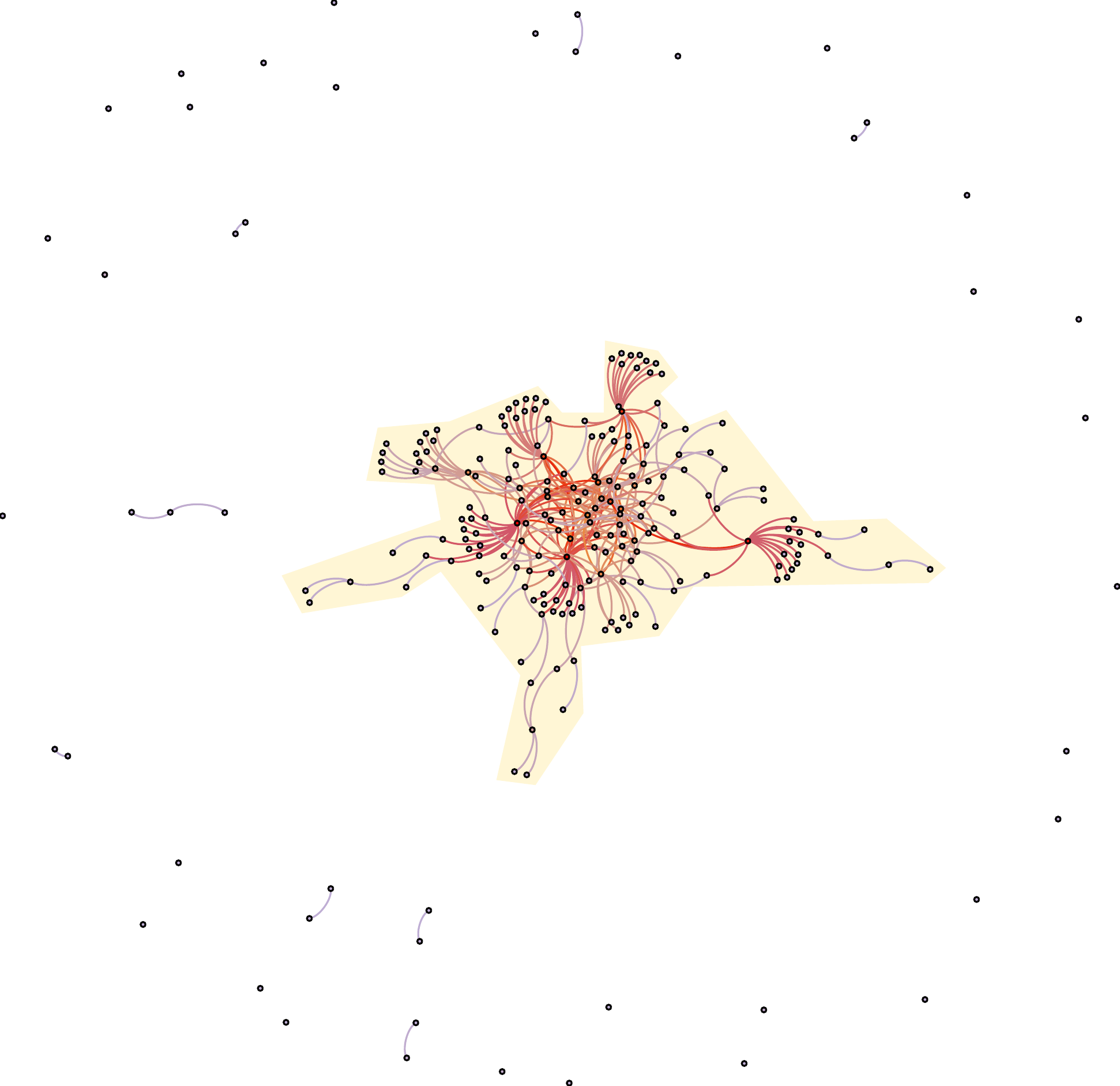} }
\subfigure[\textsc{Netbeans} (Jun. 2006) 246 nodes, 513 links  \label{fig:net_netbeans}]{
    \includegraphics[width=0.43\textwidth]{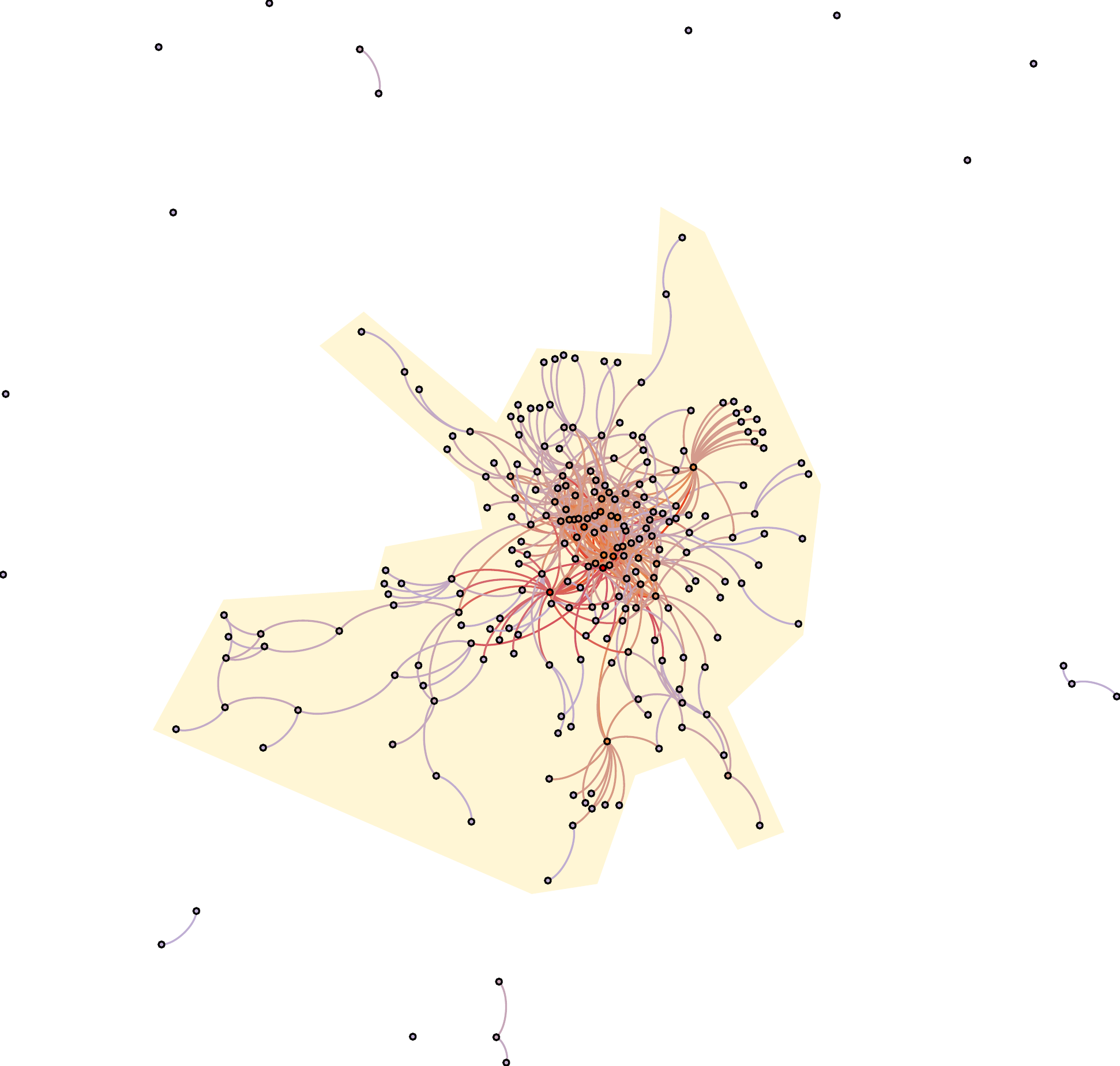} }
\\
\subfigure[\textsc{Firefox} (Oct. 2003) 241 nodes, 184 links  \label{fig:net_firefox}]{
    \includegraphics[width=0.45\textwidth]{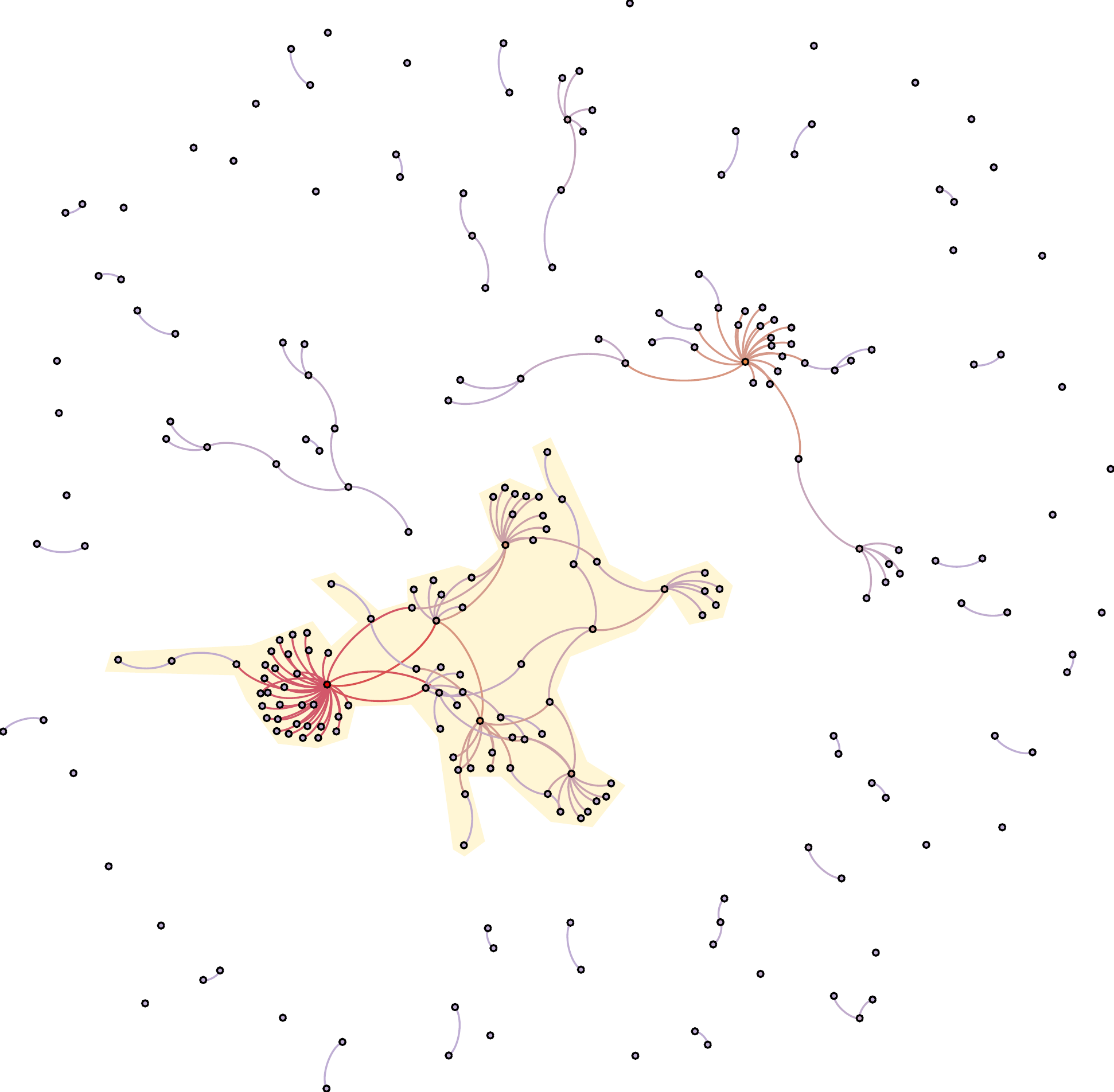} }
\subfigure[\textsc{Thunderbird} (Nov. 2004) 245 nodes, 170 links  \label{fig:net_thunderbird}]{
    \includegraphics[width=0.45\textwidth]{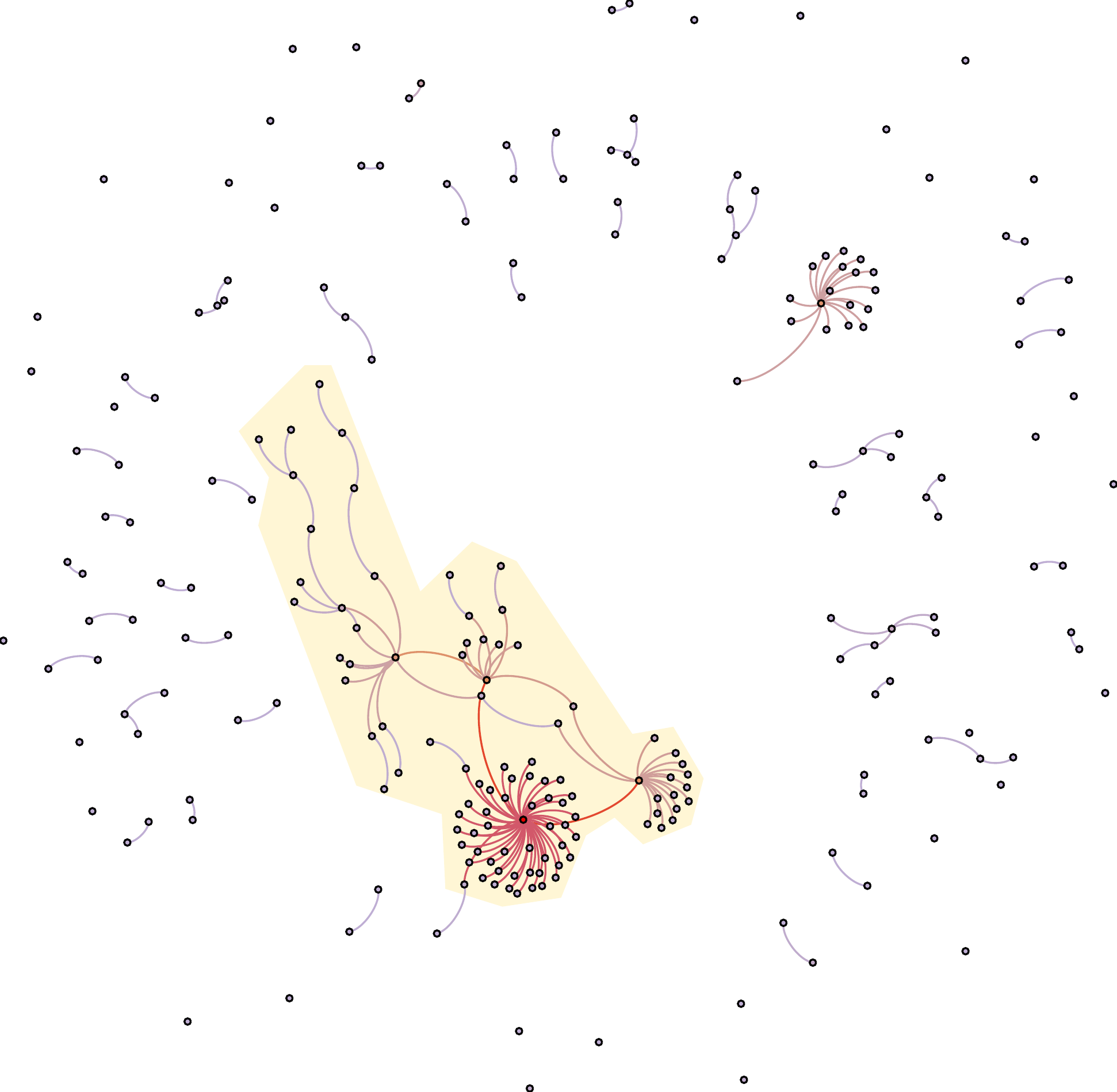} }

\caption{Four monthly collaboration networks representing the communities of \textsc{Eclipse}, \textsc{Netbeans}, \textsc{Firefox} and \textsc{Thunderbird}. Although the networks are of similar size, the different topological structures indicate that these communities differ largely in terms of social organization. The yellow shaded area represents the network's largest connected component (LCC).\label{fig:nets}}
\end{center}
\end{figure*}

In large-scale, sparse social networks usually not all nodes have a link to the rest of the network, i.e. some parts can be isolated. Thus, in addition to connected parts (components) of the network, a number of disconnected components exist.
Several network measures, including eigenvector centrality, are  not well defined for networks with different connected components.
To overcome this problem, we restrict our analysis to the so-called \emph{largest connected component} (LCC) of the monthly collaboration networks.
We find that the fraction of nodes in the LCC was high: For \textsc{Eclipse}, an average fraction of $0.78$ of all users in the monthly collaboration network belong to the LCC, for \textsc{NetBeans} the average fraction is $0.96$, for \textsc{Mozilla Thunderbird} $0.53$ and for \textsc{Mozilla Firefox} $0.58$.
Moreover, we verified that the largest size of the remaining components was insignificant when compared to the size of the LCC.
To illustrate our approach, in Figure \ref{fig:nets} we show the components of a monthly collaboration network for each of the four projects studied in our analysis.
In each of these networks of  comparable size the LCC is highlighted.
Structural differences between these networks indicate significant variations in the social organization of the four projects.
  
\section{User Centrality and Bug Report Quality\label{sec:social}}

In this section we apply the methods introduced in section \ref{sec:methods} to address research question {\textbf RQ1}, specifically:

\emph{Is the centrality of bug reporters in the collaboration network related to the quality of the submitted bug reports?}

A positive answer to this question could serve as a foundation for the development of automated bug classification schemes that are based on methods from social network analysis.
We investigate this question for four major OSS projects that adopt the \textsc{Bugzilla} bug tracking system: \textsc{Eclipse}, \textsc{Netbeans}, \textsc{Mozilla Firefox} and \textsc{Mozilla Thunderbird}.
Using the data set described in section \ref{sec:data}, we analyze the history of all bugs that were eventually marked as \emph{resolved}, along with the corresponding resolution categories.
As emphasized before, the \emph{resolved} bugs are the ones for which the bug report processing was completed (see section \ref{sec:data} for details).
The resolution categories are: \emph{FIXED}, \emph{DUPLICATE}, \emph{INVALID}, \emph{WONTFIX} and \emph{INCOMPLETE}.
In addition, we consider bugs to fall in the category \emph{INCOMPLETE}, if a bug report had this status at some point in its history, independently of the final resolution category.
According to the bug handling guidelines of the respective communities, bug reports will only be marked as such if the reporter failed to include the required additional information within a certain period of time.
Some basic statistics about the total and relative number of bugs falling in the different categories are given in Table \ref{tab:TableStatistics}.

In line with our research question, we first hypothesize that the submission of ``helpful'' bug reports - those that eventually result in a bug fix - increases the centrality of the bug reporter, i.e.

{ \bfseries H1:} {\em The centrality of users increase after the submission of bug reports that eventually result in a bug fix.}

Complementary to {\textbf H1 } we can furthermore hypothesize:

{ \bfseries H2:} {\em The centrality of users decrease after the submission of bug reports that are eventually identified as duplicate or invalid.}

While these two hypotheses address the relation between the submission of helpful or duplicate bug reports and \emph{subsequent changes} of the users' centrality in the community, it is also reasonable to consider an inverse dependence:
The users' centrality at the time when a bug is reported can possibly influence their ability to contribute helpful bug reports.
A better knowledge of bug handling procedures that results from a higher centrality in the community may for instance help to prevent duplicate bug reports.
In our third hypothesis - which is also the basis for our prediction method - we thus propose that the centrality of bug reporters is indicative for the outcome of the bug handling process.

{ \bfseries H3:} { \em The centrality of a bug reporter in the monthly collaboration network preceding the time of the report is indicative for the eventual outcome of the bug handling process.}

We would like to emphasize that one can imagine different mechanisms, both at the level of the user and the community that are compatible with these hypotheses.
As mentioned above, the users' centrality in the network is likely to be correlated with the level of contribution as well as the knowledge and experience of contributors.
These factors are likely to influence the quality of bug reports submitted by a user.
Furthermore, being central in the community can influence the attention received by other users, thus increasing the chance of bug reports being taken seriously, prioritized and eventually fixed.
 
\begin{table}[p]
\centering 
\caption{\label{tab:TableTests}Comparison of eigenvector centrality distributions for the five bug resolution categories considered in our analysis. In each row we present the hypothesis being tested, the corresponding distributions involved (e.g. $FIX_1\sim FIX_2$), the alternative hypothesis (i.e. $>$,$<$,$\neq$), its respective $p$-value (we indicate with (*) when we accept the alternative hypothesis) and the sample size of each distribution (i.e. number of bugs). More details in section \ref{sec:analysis}.}
\rotatebox{90}{
\scalebox{0.7}{
\begin{tabular}{|c|c|p{4cm}|p{4cm}|p{4cm}|p{4cm}|}
Hypothesis & Comparison of Distrib. & \textsc{Firefox} & \textsc{Thunderbird} & \textsc{Eclipse} & \textsc{Netbeans} \\
\hline
 \rowcolor{Gray} &&$<$, $p=0.0026$, (*)&$>$, $p=0.0351$, (*)&$\neq$, $p=0.1453$&$\neq$, $p=0.6435$ \\
\rowcolor{Gray}\multirow{-2}{*}{H1}&\multirow{-2}{*}{$FIX_1 \sim FIX_2$}&(5847,  6140)&(2139,  2377)&(66208, 69026)&(13930, 14668) \\ \hline
&&$>$, $p=0.0349$, (*)&$>$, $p<2.22e-16$, (*)& $>$, $p<2.22e-16$, (*)&$>$, $p<2.22e-16$, (*) \\
\multirow{-2}{*}{H2}&\multirow{-2}{*}{$DUP_1 \sim DUP_2$}&(6799,  8697)&( 973,  3027)&(17600, 22215)&(3984,  5470) \\
\rowcolor{Gray} &&$\neq$, $p=0.7268$&$>$, $p=0.0449$, (*)&$\neq$, $p=0.8489$&$\neq$, $p=0.1266$ \\
\rowcolor{Gray} \multirow{-2}{*}{H2}&\multirow{-2}{*}{$INV_1 \sim INV_2$}&(1321,  1394)&(242,   297)&(5313,  5958)&(1906,  2066) \\ \hline
&&$>$, $p=1.81e-10$, (*)&$>$, $p=1.58e-06$, (*)&$<$, $p<2.22e-16$, (*)&$>$, $p<2.22e-16$, (*) \\
\multirow{-2}{*}{H3}&\multirow{-2}{*}{$FIX_1 \sim WOF_1$}&(5847,  1022)&(2139,   106)&(66208,  7769)&(13930,  2847) \\
\rowcolor{Gray} &&$>$, $p<2.22e-16$, (*)&$>$, $p<2.22e-16$, (*)&$<$, $p<2.22e-16$, (*)&$>$, $p<2.22e-16$, (*) \\
\rowcolor{Gray} \multirow{-2}{*}{H3}&\multirow{-2}{*}{$FIX_1 \sim DUP_1$}&(5847,  6799)&(2139,   973)&(66208, 17600)&(13930,  3984) \\
&&$>$, $p<2.22e-16$, (*)&$>$, $p=4.93e-10$, (*)&$<$, $p<2.22e-16$, (*)&$>$, $p<2.22e-16$, (*) \\
\multirow{-2}{*}{H3}&\multirow{-2}{*}{$FIX_1 \sim INV_1$}&(5847,  1321)&(2139,   242)&(66208,  5313)&(13930,  1906) \\
\rowcolor{Gray} &&$>$, $p<2.22e-16$, (*)&$>$, $p<2.22e-16$, (*)&(-)(-)&$>$, $p<2.22e-16$, (*) \\
\rowcolor{Gray} \multirow{-2}{*}{H3}&\multirow{-2}{*}{$FIX_1 \sim INC_1$}&(5847,   587)&(2139,   159)&(66208,     0)&(13930,   661)
\end{tabular}}
}
\end{table}

\subsection{Analysis\label{sec:analysis}}

We test hypotheses {\textbf H1}, {\textbf H2} and {\textbf H3} in the following way:
We first categorize all bug reports that were eventually resolved according to their final resolution.
As described in section \ref{sec:nets}, we then extract the collaboration networks in the month preceding and following the time of the bug report and compute the eigenvector centrality of bug reporters in both networks.
By this, we obtain five distributions of centralities of bug reporters in the monthly collaboration network \emph{preceding} the time of the bug report for the bug categories \emph{FIXED}, \emph{DUPLICATE}, \emph{INVALID}, \emph{WONTFIX} and \emph{INCOMPLETE}.
We denote these as $FIX_1$, $DUP_1$, $INV_1$, $WOF_1$ and $INC_1$ respectively.
Similarly, we extract the distributions of eigenvector centralities of bug reporters in the month \emph{after} the bug report and denote these as $FIX_2$, $DUP_2$, $INV_2$, $WOF_2$ and $INC_2$.
We would like to emphasize that - out of the quantitative measures introduced in section \ref{sec:measures} - in this section we only use eigenvector centrality to quantify the position of bug reporters.
However, for the classifier proposed in the next section we use a more comprehensive set consisting of additional topological measures for centrality, coreness, degree and membership in the LCC.

In order to compare the different eigenvector centrality distributions of bug reporters described above, we apply a \emph{Wilcoxon-Mann-Whitney} test \cite{WMWtest1999}.
For two samples $S_A$ and $S_B$ drawn from two distributions $F_A$ and $F_B$ with $F_A(x) = F_B(x-\alpha)$, the \emph{Wilcoxon-Mann-Whitney} infers the stochastic ordering of the distributions, i.e. whether the shift parameter $\alpha$ is likely to be larger than zero (i.e. $F_A > F_B$) or smaller than zero (i.e. $F_A < F_B$).
Based on the null hypothesis that $\alpha = 0$ (i.e. $F_A \sim F_B$) the test is executed either with the one-sided alternative hypotheses $F_A > F_B$ or $F_A < F_B$, or with a two-sided alternative hypothesis $F_A \neq F_B$.
For each of the three alternative hypotheses, the test yields a $p$-value which - if it is below a given significance threshold - is used to reject the null hypothesis in favor of the alternative hypothesis.
If none of the $p$-values for one of the alternative hypotheses is below the significance threshold, one cannot reject the \emph{null hypothesis} that both samples $S_A$ and $S_B$ are in fact drawn from the same distribution, i.e. $F_A \sim F_B$.

We now test {\textbf H1} by applying the methodology described above to the two samples $FIX_1$ and $FIX_2$, i.e. we test whether there is an increase in the eigenvector centralities of users after the report of a bug that is eventually fixed.
The null hypothesis {\textbf H0} related to {\textbf H1} is that the samples $FIX_1$ and $FIX_2$ are drawn from the \emph{same distribution}, i.e. $FIX_1 \sim FIX_2$ or - in other words - the eigenvector centrality of users reporting helpful bugs \emph{does not change} after the time of the report.
We reject the null hypothesis and accept hypothesis \textbf{H1} if the $p$-value for $FIX_1 < FIX_2$ is below a significance threshold of $0.05$.
The resulting $p$-values for the comparison of the distributions $FIX_1$ and $FIX_2$ are given in Table \ref{tab:TableTests}.
One observes that for the projects \textsc{Eclipse} and \textsc{NetBeans} one cannot reject the null hypothesis that eigenvector centralities of users \emph{do not change} after the submission of bug reports that result in a bug fix.
However, for \textsc{Mozilla Firefox} \emph{there is a significant increase} in the eigenvector centralities of users reporting bugs that are eventually fixed.
Interestingly, for \textsc{Mozilla Thunderbird} \emph{we also reject the null hypothesis but instead find a significant decrease of eigenvector centrality}.

Similar to {\textbf H1}, we test hypothesis {\textbf H2} by applying a \emph{Wilcoxon-Mann-Whitney} test on the samples $DUP_1$, $INV_1$, $DUP_2$ and $INV_2$, i.e. we compare the eigenvector centrality distributions of bug reporters submitting duplicate or invalid bug reports \emph{before} and \emph{after} the time of the submission.
The results shown in Table \ref{tab:TableTests} provide strong evidence for hypothesis { \textbf H2} regarding bugs that are eventually identified as duplicates.
In fact, the null hypothesis that $DUP_1$ and $DUP_2$ are drawn from the same distribution can be rejected in favor of the alternative hypothesis $DUP_1 > DUP_2$ for all of the studied projects.
For the case of bugs that are eventually identified as invalid, we cannot reject the null hypothesis for the projects \textsc{Firefox}, \textsc{Eclipse} and \textsc{NetBeans}.
For the project \textsc{Thunderbird} the null hypothesis can be rejected in favor of hypothesis {\textbf H2}.

Finally, we test hypothesis {\textbf H3} by comparing the distribution $FIX_1$ to the distributions $WOF_1$, $DUP_1$, $INV_1$ and $INC_1$, i.e. we check whether the centralities of users reporting bugs that are eventually fixed are - on average - different than of those reporting bugs that fall in other categories.
The results of our analysis are shown in Table \ref{tab:TableTests}.
We find strong evidence for hypothesis {\textbf H3} when comparing $FIX_1$ to either $WOF_1$, $DUP_1$, $INV_1$ or $INC_1$.
In the projects \textsc{Firefox}, \textsc{Thunderbird} and \textsc{Netbeans} we particularly find that the centrality of users reporting bugs that are eventually fixed is significantly larger.
Interestingly, the opposite relation holds for the project \emph{Eclipse}, i.e. here the centrality of users reporting bugs that are eventually fixed is significantly smaller.

In summary, our analysis validates that there is a statistically significant relation between the centrality of a bug reporter and the outcome of bug handling processes.
We particularly emphasize that our analysis supports the hypothesis that the centrality in the collaboration network during the month preceding the bug report is indicative for the outcome of the bug handling process.
In the following section, we make use of this finding to develop a prediction method that can e.g. be applied in (semi-)automatic bug report prioritization strategies.
By this, we show that a quantitative analysis of social structures in OSS communities can assist in bug triaging.
While in the next section we exclusively focus on the use of a set measures of \emph{social embeddedness}, we would like to highlight that a combination of these measures with existing methods is likely to further improve the classification mechanism.

\section{Classification of Bugs with Social Network Analysis\label{sec:classify}}


Based on the observed relations between the bug reporters' centrality and bug report quality presented in section \ref{sec:social}, we now 
address research question {\bfseries R}Q2, specifically:

\emph{Can quantitative measures for the position of bug reporters be used to predict which bug reports refer to valid bugs?}

The goal is to develop a practical method that makes use of topological measures for the position of bug reporters in the collaboration network.
In order to facilitate the bug triaging process, we particularly aim at predicting whether a bug report is likely to be either \emph{Valid} or \emph{Faulty}.
As \emph{Valid} bug reports, we consider all bug reports that have a final status of \emph{FIXED} or \emph{WONTFIX}.
Conversely - and in line with the semantics of bug categories provided in section \ref{sec:data} - we consider all bug reports as \emph{Faulty} that have a final status of \emph{DUPLICATE}, \emph{INVALID} or \emph{INCOMPLETE}.

\begin{figure}[htpb]
\centering
\subfigure[Classifier based solely on LCC membership\label{fig:lccbox}]{
    \includegraphics[width=0.46\textwidth]{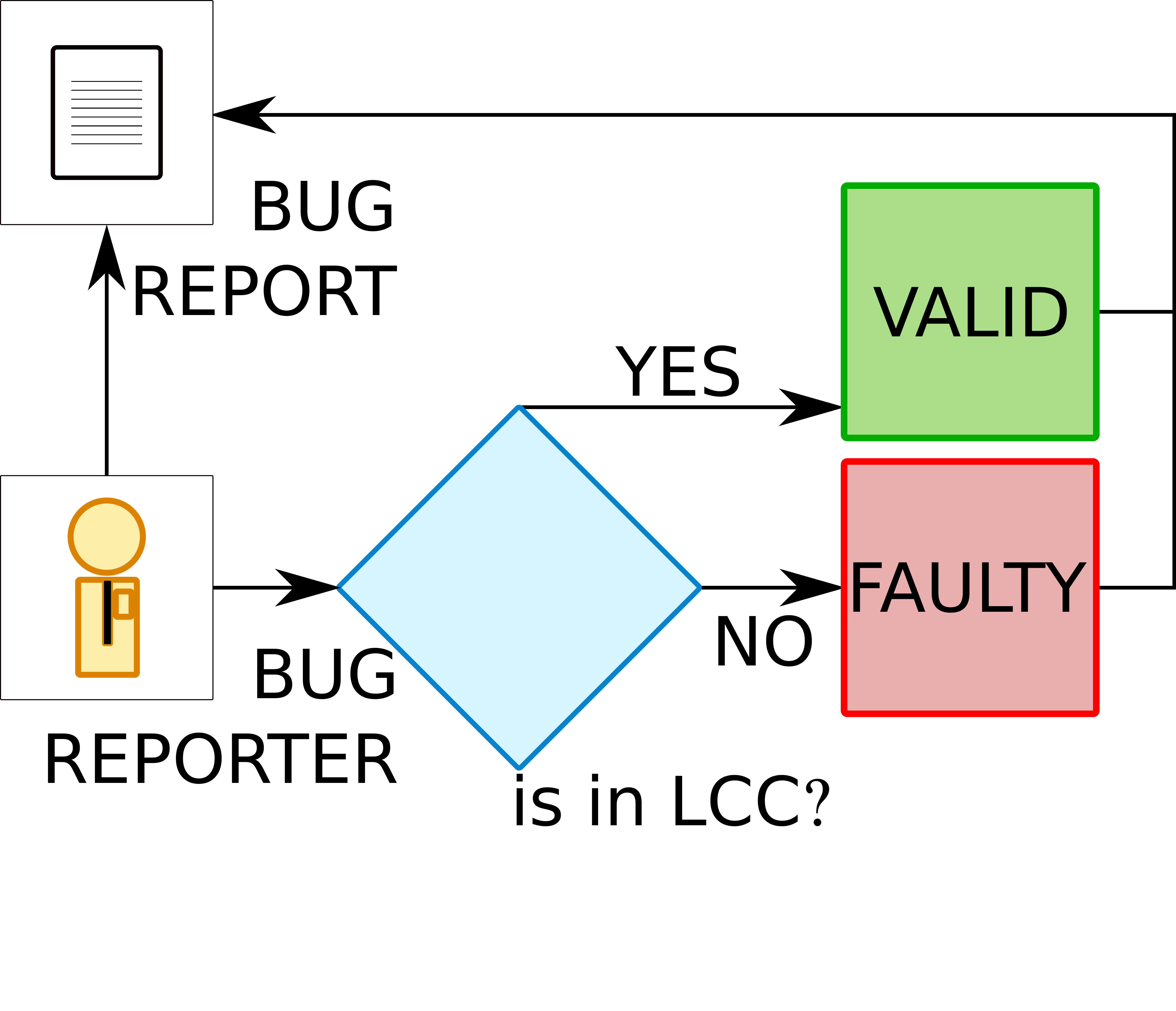} }
\subfigure[Thresholding on eigenvector centr. of LCC nodes\label{fig:lccebox}]{
    \includegraphics[width=0.46\textwidth]{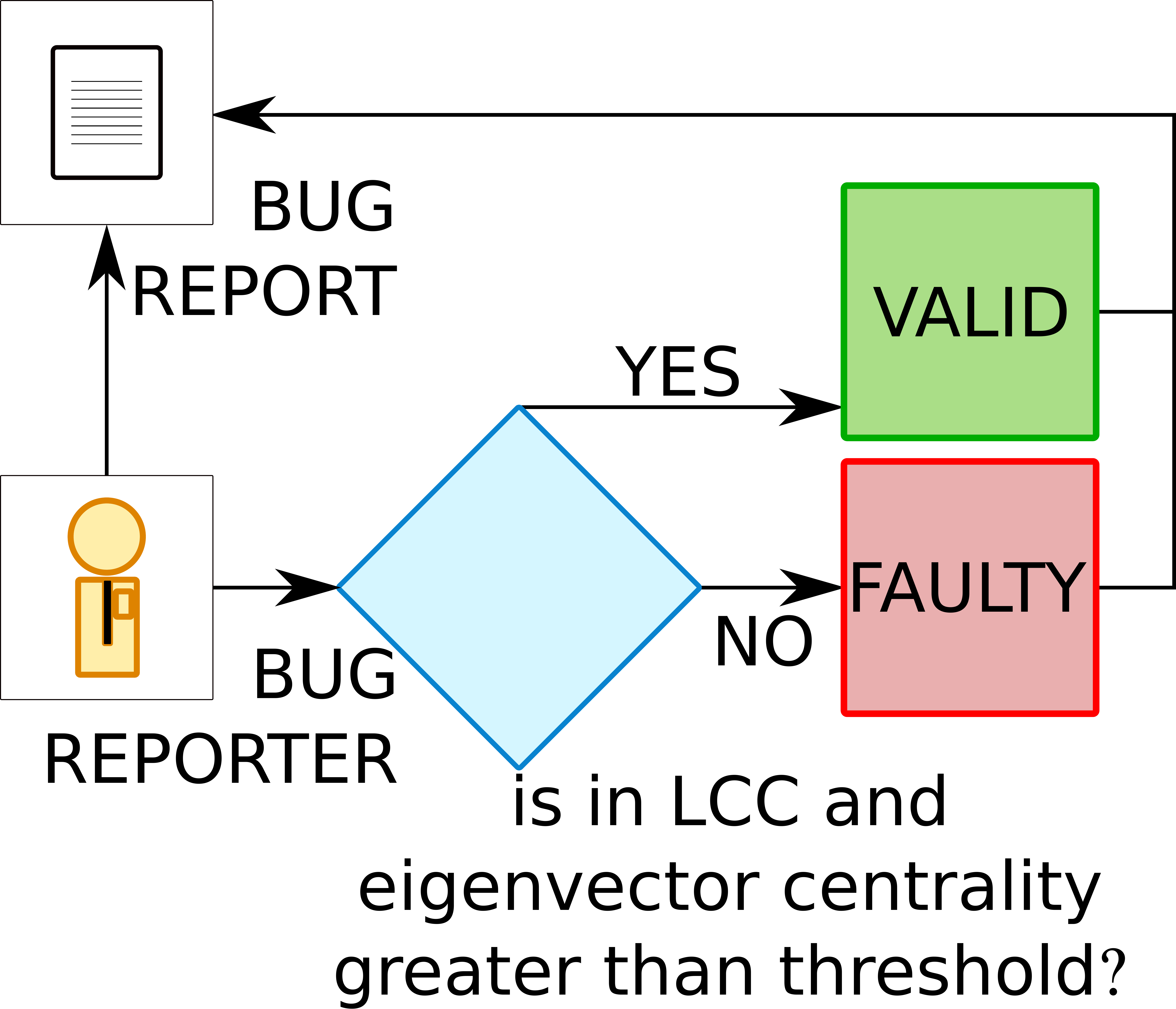} }\\
\subfigure[SVM Classifier\label{fig:svmbox}]{
    \includegraphics[width=0.64\textwidth]{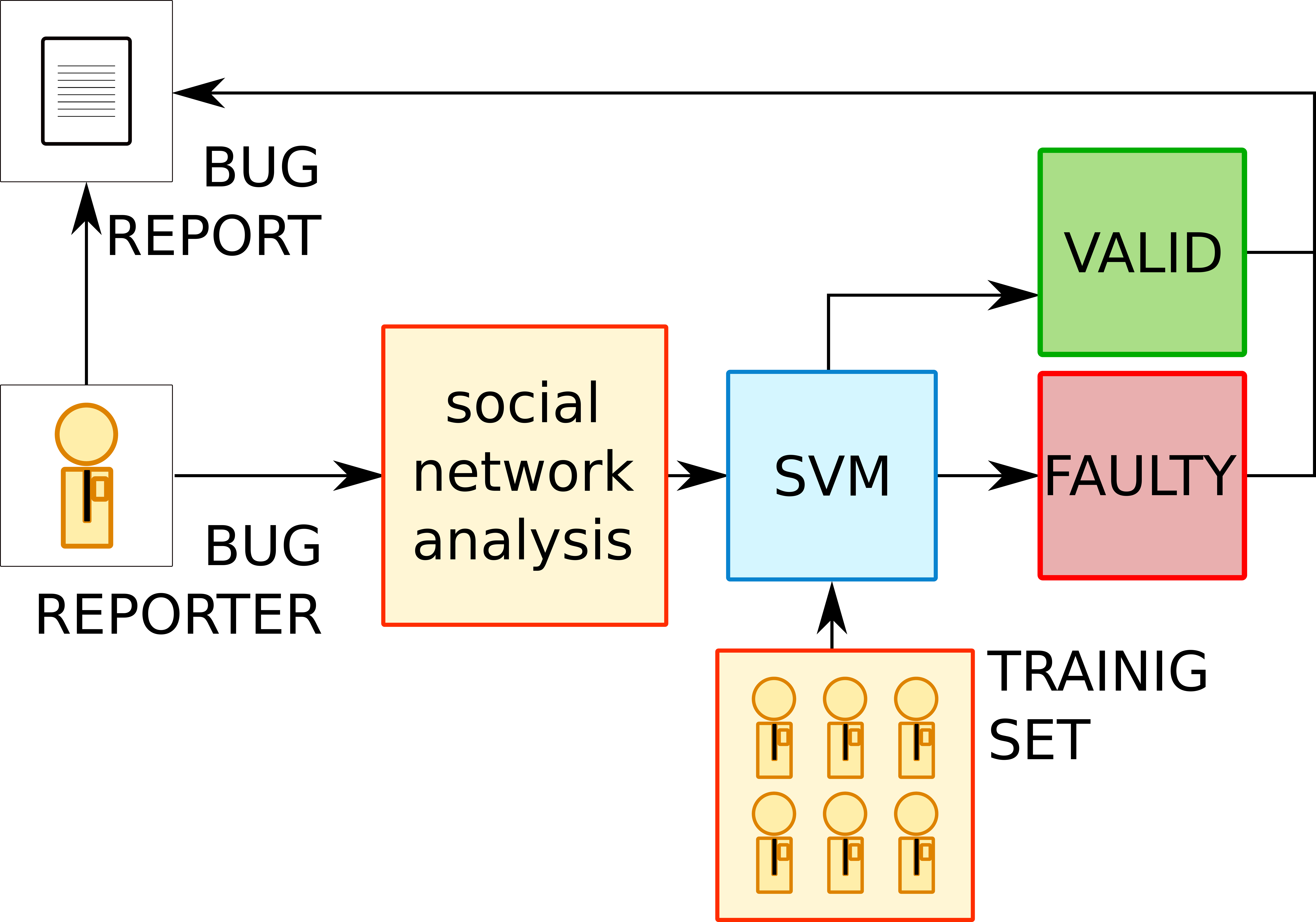} }
\caption{Graphical illustration of the three classifiers described in section \ref{sec:classify}. When bug reporters submit reports, we immediately quantify the nine measures that express their social embeddedness as described in the text. These are used as input to the classifier, which will then predict if bug reports are valid or faulty. For the case of the SVM classifier, we separate $5.0\%$ of the samples to be used as training data.\label{fig:svm}}
\end{figure}

The task for our classifier is to predict whether a given bug report is \emph{Valid} or \emph{Faulty}, based on a set of features that are comprised of different quantitative measures for the position of bug reporters in the collaboration network.
In order to highlight the predictive power gained by the inclusion of further measures, we start with a very simple classifier which only considers the presence of a bug reporter in the network's largest connected component (LCC).
We then incrementally add a prediction that is based on a threshold of eigenvector centrality as well as - eventually - a support vector machine that makes use of the following set of nine topological measures calculated at the level of a node: presence in the LCC, eigenvector, betweenness, and closeness centrality, local clustering coefficient, coreness, as well as in-, out- and total degree.
Illustrative overviews of the three different classification schemes are provided in Figures \ref{fig:lccbox} - \ref{fig:svmbox}.
For each of the obtained classifiers, we evaluate its predictive power in terms of \emph{precision}, \emph{recall} and the corresponding \emph{F}-score (i.e. equally weighted precision and recall) \cite{info1986, Hooimeijer2007}.
In order to enable the reader to correctly interpret the predictive power based on the obtained precision and recall values, in the first line of Table \ref{tab:TablePrecRecall} we indicate the actual fraction of \emph{Valid} bug reports in our data set for each of the considered projects.

\begin{table}[htpb]  
\centering
\caption{Percentages of bug reporters that are in the LCC of the
  social network in the month preceding the report. The percentages
  given were calculated for each of the resolution categories
  (e.g. for \textsc{Firefox}, from those that reported bugs resolved as FIXED: 53.9\% were in the LCC while 46.1\% were not). }
\scalebox{0.8}{
\begin{tabular}{|l|r|r|r|r|r|r|}
\label{tab:TableLCCFracs}
                                & \textsc{Firefox}           & \textsc{Thunderbird}           & \textsc{Eclipse}       & \textsc{Netbeans} \\
\hline
 \rowcolor{Gray}    FIX         & $53.9 \%$         & $47.4 \%$         & $64.0 \%$     & $65.0 \%$    \\
                    DUP         & $28.0 \%$         & $9.4 \%$          & $62.4 \%$     & $42.7 \%$   \\
 \rowcolor{Gray}    INV         & $11.2 \%$         & $8.6 \%$          & $42.2 \%$     & $46.7 \%$     \\
                    WOF         & $37.7 \%$         & $18.2 \%$         & $52.9 \%$     & $51.6 \%$   \\
 \rowcolor{Gray}    INC         & $4.1 \%$          & $4.7 \%$          & -             & $26.6 \%$    \\
 \hline
  Valid                         & $50.6 \%$         & $44.1 \%$         & $62.6  \%$    & $62.2 \%$    \\
  \rowcolor{Gray}                  Faulty      & $17.2 \%$         & $8.3 \%$          & $56.1 \%$     & $41.2 \%$
\end{tabular}}
\end{table}

We first consider a simple prediction method which considers a bug report to be valid whenever the bug reporter is in the LCC of the collaboration network in the month preceding the submission of the bug report.
The basis for this prediction is provided in Table \ref{tab:TableLCCFracs}, which lists the fraction of bug reporters belonging to the LCC of the network individually for each of the different bug categories.
In the two bottom rows, we furthermore provide the same values for the aggregated sets of \emph{Valid} and \emph{Faulty} bugs.
For \textsc{Mozilla Firefox} and \textsc{Mozilla Thunderbird} one observes a significant difference between these two categories, i.e. the fraction of reporters of \emph{Valid} bugs that are in the LCC is significantly higher than the fraction of reporters of \emph{Faulty} bugs.
For \textsc{Eclipse} and \textsc{NetBeans} the effect is less pronounced.
Table \ref{tab:TablePrecRecall} (i.e. (LCC) rows) shows the precision, recall and $F$-score of a classifier that is solely based on LCC membership.
When comparing to the real proportion of \emph{VALID} bug reports, this predictor clearly performs better than a null model of randomly sampling bug reports.
Due to the stronger effect of LCC membership, the performance is clearly better for \textsc{Mozilla Firefox} and \textsc{Mozilla Thunderbird}, which at the same time are the projects with the smallest proportion of \emph{VALID} bug reports.

\begin{table}[ht]  
\centering
\caption{Precision ($p$), recall ($r$) and $F$-score of filtering valid bug reports based only on measures of social embeddedness.}
\scalebox{0.8}{
\begin{tabular}{|l|r|r|r|r|r|r|}
\label{tab:TablePrecRecall}
                                    & \textsc{Firefox}        & \textsc{Thunderbird}  & \textsc{Eclipse}    & \textsc{Netbeans}   \\
\hline
 \rowcolor{Gray}    Valid   & $21.0 \%$      & $23.3 \%$    & $74.3 \%$  & $62.4 \%$ \\
\hline
                    $p$ (LCC)       & $44.1 \%$      & $62.1 \%$    & $76.3 \%$  & $71.9 \%$ \\
 \rowcolor{Gray}    $r$ (LCC)       & $50.9 \%$      & $44.5 \%$    & $62.6 \%$  & $62.4 \%$ \\
                    $F$ (LCC)       & 0.47           & 0.52         & 0.69       & 0.67      \\
\hline
 \rowcolor{Gray}    $p$ (evcent)    & $60.4 \%$      & $68.6 \%$    & $76.3\%$   & $76.7\%$  \\
                    $r$ (evcent)    & $30.5 \%$      & $ 5.4 \%$    & $62.6\%$   & $38.8\%$  \\
 \rowcolor{Gray}    $F$ (evcent)    & $0.41$         & $0.10$       & $0.69$     & $0.52$      \\

\hline

                    $p$ (SVM)       & $82.5 \%$      & $90.3 \%$    & $88.7 \%$  & $78.9 \%$ \\
 \rowcolor{Gray}    $r$ (SVM)       & $44.5 \%$      & $38.9 \%$    & $91.0 \%$  & $87.0 \%$ \\
                    $F$ (SVM)       & $0.58$         & $0.54$       & $0.89$     & $0.83$
\end{tabular}}
\end{table}

As the next measure we add to the classifier the eigenvector centrality of bug reporters.
This classifier will mark bug reports as \emph{VALID} if the reporting users is part of the LCC and if their respective eigenvector centrality scores are above a precentile threshold that is tuned for each community individually.
The results shown in Table \ref{tab:TablePrecRecall} (i.e. (evcent) rows) indicate that - compared to a classification based on mere LCC membership - the inclusion of eigenvector centrality increases the precision while generally decreasing recall and $F$-score.
Due to the negative relation between eigenvector centrality and bug report quality found for \textsc{Mozilla Thunderbird}, the drop in the $F$-score is particularly pronounced for this project.

Our next and final step towards a practical tool is a) the use of a support vector machine (SVM) \cite{statisticallearning2005} for the prediction of \emph{valid} bug reports and b) the use of the full set of nine topological measures.
In order to eliminate the risk of overfitting the data, we use a training set that is composed of only $5.0\%$ of all available samples.
The nine measures we consider as input features are: \emph{LCC membership}, \emph{eigenvector centrality}, \emph{betweenness centrality}, \emph{total degree}, \emph{in-degree}, \emph{out-degree}, \emph{closeness centrality}, \emph{clustering coefficient} and \emph{k-coreness}.
We present the results of the SVM classifier in Table \ref{tab:TablePrecRecall} (i.e. (SVM) rows).
For \textsc{Mozilla Firefox} and \textsc{Mozilla Thunderbird} we obtain precision values of $82.5$ and $90.3$ as well as $F$-scores of $0.58$ and $0.54$ respectively.
In both of these projects the fraction of \emph{Valid} bug reports is comparably small (with $21 \%$ and $23.3 \%$ respectively).

The fraction of \emph{Valid} bugs in the \textsc{Eclipse} and \textsc{NetBeans} projects is significantly higher.
We hypothesize that this is due to more stringent bug reporting procedures and a higher technical proficiency of users which is related to the fact that both projects target a user community that mainly consists of developers.
For \textsc{Eclipse} and \textsc{NetBeans} our classifier obtains a precision of $88.7 \%$ and $78.9 \%$ with $F$-scores of $0.89$ and $0.83$ respectively.
Since the majority of bug reports in these two projects are \emph{Valid}, we propose to use the classifier to identify the minority of \emph{Faulty} bug reports instead.
In Table \ref{tab:TablePrecRecallFaulty}, we show the corresponding results for all four projects.
In this setting, our classifier achieves $F$-scores of $0.92$ and $0.91$  and a precision of $86.9 \%$ and $84.9 \%$ for \textsc{Mozilla Firefox} and \textsc{Mozilla Thunderbird} respectively.
For the projects \textsc{Eclipse} and \textsc{NetBeans} we obtain a precision of $73.6 \%$ and $73.1 \%$ and $F$-scores of $0.69$ and $0.67$ respectively.

\begin{table}[ht]
\centering
\caption{Precision ($p$), recall ($r$) and $F$-score of filtering faulty bug reports based only on measures of social embeddedness.}
\scalebox{0.8}{
\begin{tabular}{|l|r|r|r|r|r|r|}
\label{tab:TablePrecRecallFaulty}
                              & \textsc{Firefox}& \textsc{Thunderbird} & \textsc{Eclipse}       & \textsc{Netbeans} \\
\hline
\rowcolor{Gray}       Faulty  & $79.0\%$        & $76.7\%$             & $25.7 \%$              & $37.6 \%$ \\
\hline

                    $p$ (SVM) & $86.9\%$        & $84.9\%$             & $ 73.6\%$              & $73.1 \%$ \\
\rowcolor{Gray}     $r$ (SVM) & $97.3\%$       & $98.2\%$             & $ 64.0\%$               & $61.8 \%$ \\
                    $F$ (SVM) & $0.92$          & $0.91$              & $0.69$               & $0.67$ \\
\end{tabular}}
\end{table}

\section{Discussion, Threats to Validity and Implications for Future Work}
\label{sec:threats}


Prior to concluding our article, we discuss a number of limitations of our analysis as well as resulting threats to validity.
As described in section \ref{sec:methods}, all our findings are based on interactions recorded in the \textsc{Bugzilla} installation of the projects \textsc{Mozilla Firefox}, \textsc{Mozilla Thunderbird}, \textsc{Eclipse} and \textsc{Netbeans}.
Clearly, a significant threat to the applicability of our approach for general collaborative software engineering is that we were mainly focused on these four OSS communities.
However, we argue that these particular projects represent communities with different levels of heterogeneity with respect to the level of contributions, commitment, technical proficiency and commercial influence by companies.
In particular, the communities of \textsc{Mozilla Firefox} and \textsc{Mozilla Thunderbird} target a rather general audience without particular technical proficiency, while \textsc{Eclipse} and \textsc{Netbeans} are more focused on software developers.
As such, our particular choice of communities may be considered as covering different ends of the spectrum of technical proficiency of users.
Our analysis shows that, even for such diverse projects, machine learning techniques based on quantitative measures of social embeddedness yield high accuracy results when predicting bug report quality. 
Therefore our contribution can be seen as a proof of concept case study.
Nevertheless, we are currently collecting and analyzing data as well as qualitative insights on the social organization of a number of additional communities in order to generalize our results.
 
Although our analysis focuses on the \textsc{Bugzilla} communities of OSS projects, our methodology is - in general - not limited to these.
Any issue tracking system which records time-stamped direct interactions between its users can be used to extract evolving collaboration networks and thus to compute quantitative measures for social embeddedness.
However, whether these measures can be used for highly accurate, automated bug categorization in settings other than the ones studied in this paper (like e.g. commercial software production or collaborations in smaller or less diverse teams) requires further studies and is beyond the scope of our work.

While we have presented a set of quantitative results regarding the relation between the network position of bug reporters and the outcome of bug report processing, it is unclear what are the exact social mechanisms at work.
In order to gain a better insight into this question, we have created a survey that was sent to the community managers of the projects considered in this case study.
Indeed, in their replies the community managers of \textsc{Eclipse} and \textsc{NetBeans} confirmed that such a relation may exist.
Specifically, we received feedback indicating that for the \textsc{NetBeans} community ``one of the criteria developers use while choosing bugs for fixing is reproducible case and/or reputation of the reporter''.
Similarly, for the \textsc{Eclipse} project community managers confirmed that ``a committer is often times more likely to spend triage time on a bug from somebody with a known reputation for quality''.
Unfortunately, we did not receive any feedback to our survey for the communities of \textsc{Mozilla Firefox} and \textsc{Mozilla Thunderbird}.

For the network measures studied in this paper, we only used the direct dyadic relations \emph{CC} (i.e. users subscribing to receive information about future updates on bug reports) and \emph{Assign} (i.e. users assigning the task of handling a bug to another one).
While these recorded interactions are clearly associated with users knowing about and interacting with each other, the resulting network must clearly be seen as a mere proxy for the actual social organization of a community.
In particular, in our study of network measures we did not consider further relations that may be extracted for instance from the sequence of comments on a bug.
The reason for not considering these is the lower fidelity with respect to whether an extracted relation is really associated with direct communication or collaboration.
Furthermore, in our study we so far did not use further potential data sources, like mailing lists or threaded forum communication that could be used to augment our network perspective in a subsequent analysis.

Another remark related to the measures of social embeddedness adopted in our analysis is that they can be quantified right away after a bug report is submitted.
As we show in the paper, this works well for OSS communities that have accumulated enough samples to apply machine learning techniques.
Therefore the extension of this methodology to newly born communities remains a challenge.

A possible reason of concern is the fact that we use a fixed-size window of $30$ days to construct the networks used in our analysis. 
Although we have obtained high accuracy results for this particular choice of window size, we are further investigating whether tuning this parameter to each community independently will further increase performance.

Finally, the application of machine learning comes at the risk of overfitting data by using a too large fraction of training data.
In order to avoid this pitfall, we limited the fraction of randomly chosen training data to $5.0 \%$. 
To foster the reproducibility of our results and to facilitate the implementation of similar approaches of social awareness in practical support infrastructures, the source code of the SVM classifier (written in the \textsc{R} language) as well as the data sets studied in our analysis are available online\footnote{see \url{http://www.sg.ethz.ch/research/topics/social-se/data/}}.

\section{Conclusions}
\label{sec:conclusions}

In this paper we have studied to what extent the positions of bug reporters in the collaboration networks of four OSS communities are indicative for the quality of contributed bug reports.
We have addressed this question from the perspective of evolving complex networks that have been extracted from a comprehensive data set on $700,000$ bug reports for the projects \textsc{Mozilla Firefox}, \textsc{Mozilla Thunderbird}, \textsc{Eclipse} and \textsc{NetBeans}.
The main results of our case study on these communities are the following:


\textbf{(1)} We study the evolution of bug reporter centrality in \emph{evolving collaboration networks}, using a time resolution of $30$ days over a total period of $10$ years. For the project \textsc{Mozilla Firefox}, we are able to validate our hypothesis that the eigenvector centrality of bug reporters increases after the submission of valid bug reports (i.e. reports that refer to actual software bugs, are no duplicates and contain all necessary information). We observe the opposite relation for \textsc{Mozilla Thunderbird}.

\textbf{(2)} In all projects we were able to validate our hypothesis that there is a statistically significant decrease of eigenvector centrality following the submission of duplicate bugs.

\textbf{(3)} For the projects \textsc{Mozilla Firefox}, \textsc{Mozilla Thunderbird} and \textsc{NetBeans} we were able to validate our hypothesis that the eigenvector centrality of users reporting \emph{valid} bug reports  is significantly higher than those of users submitting \emph{faulty} bug reports. From this we conclude that the position of bug reporters in the collaboration network of OSS communities is indicative for the quality of bug reports.

\textbf{(4)} Based on this finding, we develop an automated bug report classification mechanism. We use nine topological measures at the level of bug reporters (eigenvector, betweenness and closeness centrality, k-coreness, clustering coefficient, \mbox{in-}, out- and total degree as well as membership in the largest connected component) for the prediction of whether a reported bug is \emph{valid} or \emph{faulty}. Based on a support vector machine and depending on the project considered, our automated classification achieves a precision of up to $90.3 \%$ and an $F$-score of up to $0.92$.

We would like to emphasize the fact that - although it is merely based on measures quantifying the network position of bug reporters - \emph{our proposed classification mechanism achieves a remarkably high accuracy across different communities}.
The combination of our approach with further features used in previous studies of automated bug classification is likely to further improve its accuracy.
Our case study can thus be seen as a contribution towards classification schemes that are highly accurate, yet simple enough to be of practical relevance in the design of support infrastructures.
     
\section*{Acknowledgment}
  
This work was supported by the SNF through grant CR12I1\_125298. We would like to acknowledge the contribution of Emre Sarig\"ol to the collection and preprocessing of data, and the communities of \textsc{Netbeans} and \textsc{Eclipse} for sharing their insights with us.

\bibliographystyle{acm}    
\bibliography{ref}
   
\end{document}